\documentclass[fleqn,usenatbib]{mnras}

\usepackage{newtxtext,newtxmath}
\usepackage[T1]{fontenc}
\usepackage{ae,aecompl}

\usepackage{graphicx}	% Including figure files
\usepackage{amsmath}	% Advanced maths commands
\usepackage{amssymb}	% Extra maths symbols

\newcommand\lsim{\mathrel{\rlap{\lower4pt\hbox{\hskip1pt$\sim$}}
        \raise1pt\hbox{$<$}}}
\newcommand\gsim{\mathrel{\rlap{\lower4pt\hbox{\hskip1pt$\sim$}}
        \raise1pt\hbox{$>$}}} 
\newcommand{\msun}{{\rm M_{\odot}}}

\title[Circumbinary Disc Torques]{On the orbital evolution of supermassive black hole binaries with circumbinary accretion discs}

\author[Y. Tang, A. MacFadyen and Z. Haiman]{
Yike Tang$^{1\star}$,
Andrew MacFadyen$^{1\star}$,
Zolt\'{a}n Haiman$^{2,1}$\thanks{E-mail: yt611@nyu.edu~(YT);~macfadyen@nyu.edu~(AM); zoltan@astro.columbia.edu (ZH)}
\\
$^{1}$Center for Cosmology and Particle Physics, New York University, New York, NY, USA\\
$^{2}$Department of Astrophysics, Columbia University, New York, NY, USA\\
}

\date{Accepted XXX. Received YYY; in original form ZZZ}

\pubyear{2017}

\begin{document}
\label{firstpage}
\pagerange{\pageref{firstpage}--\pageref{lastpage}}
\maketitle

% Abstract of the paper
\begin{abstract}
  Gaseous circumbinary accretion discs provide a promising mechanism to facilitate
  the mergers of supermassive black holes (SMBHs) in galactic nuclei. We measure the
  torques exerted on accreting SMBH binaries, using 2D, isothermal, moving-mesh, viscous
  hydrodynamical simulations of circumbinary accretion discs. Our computational
  domain includes the entire inner region of the circumbinary disk with the individual
  black holes (BHs) included as point masses on the grid and a sink prescription to
  model accretion onto each BH. The BHs each acquire their own well-resolved accretion
  discs ("minidiscs").  We explore a range of mass removal rates for the sink prescription
  removing gas from the central regions of the minidiscs.  We find
  that the torque exerted on the binary is primarily gravitational,
  and dominated by the gas orbiting close behind and ahead of the
  individual BHs. The torques from the distorted circumbinary disc
  farther out and from the direct accretion of angular momentum are
  sub-dominant. The torques are sensitive to the sink prescription:
  slower sinks result in more gas accumulating near the BHs and more
  negative torques, driving the binary to merger more rapidly.  For
  faster sinks, the torques are less negative, and eventually turn
  positive (for unphysically fast sinks).  When the minidiscs are
  modeled as standard $\alpha$ discs, our results are insensitive to the
  choice of sink radius.  Scaling the simulations to a binary orbital
  period $t_{\rm bin}=1$yr and background disc accretion rate
  $\dot{M}=0.3\dot{M}_{\rm Edd}$ in Eddington units, the binary
  inspirals on a timescale of $\approx 3\times10^{6}$ years,
  irrespective of the SMBH masses.  For binaries with total mass
  $\lsim 10^7~{\rm M_\odot}$, this is shorter than the inspiral time
  due to gravitational wave (GW) emission alone, implying that gas
  discs will have a significant impact on the SMBH binary population
  and can affect the GW signal for Pulsar Timing Arrays.
\end{abstract}

% Select between one and six entries from the list of approved keywords.
% Don't make up new ones.

%%%%%%%%%%%%%%%%%%%%%%%%%%%%%%%%%%%%%%%%%%%%%%%%%%

%%%%%%%%%%%%%%%%% BODY OF PAPER %%%%%%%%%%%%%%%%%%

\section{Introduction}

Binaries embedded in gas discs occur frequently in nature.  Examples
include planet+star systems with protoplanetary discs
\citep[e.g.][]{1997Icar..126..261W}, young stellar binaries
\citep[e.g.][]{ Orosz:2012Sci}, and supermassive black hole binaries
(SMBHBs) expected in many galactic nuclei
\citep[e.g.][]{Dotti:2012:rev,Mayer:2013:MBHBGasRev}. The nature of
the binary-disc interaction is of compelling interest for
understanding the long-term evolution of these systems.  In the case
of SMBHBs, in particular, a long-standing question is whether such
binaries can reach orbital separations as small as $\sim1$pc, so that
gravitational radiation would cause an inevitable
merger~\citep[e.g.][]{Begel:Blan:Rees:1980}.  Gas discs could help
facilitate orbital decay down to these separations
\citep[e.g.][]{Escala:2005}. Understanding the interaction between
SMBHBs and their gas discs is also crucial for predicting the
electromagnetic (EM) signatures of such systems
\citep[e.g.][]{TMH2012}.

Significant work has been done to understand the linear binary-disc
interaction in the limit of small mass ratio, $q\equiv M_2/M_1\ll 1$.
This so-called Type I planetary migration is facilitated by linear
spiral density waves launched at resonant disc locations
\citep[e.g.][]{GT80}. The rate and even the direction (inward or
outward) of Type I migration is sensitive to thermodynamics
\citep[e.g.][]{PM2006}.  The regime $10^{-4} \lsim q \lsim 0.05$ has
also been widely explored. The secondary opens a narrow annular gap in
the disc, resulting in so-called Type II migration
\citep[e.g.][]{1997Icar..126..261W}.  The interaction is non-linear,
generally causing inward migration on a time-scale comparable to the
viscous time-scale of the disc near the binary
\citep[e.g.][]{1986ApJ...309..846L}. Deviations from this time-scale
can, however, can be significant, especially when the mass of the
smaller BH exceeds that of the nearby disc
\citep[e.g.][]{1995MNRAS.277..758S,Ivanov99,2014ApJ...792L..10D,DK2017}.

By comparison, relatively little analogous work has been done for SMBH
binaries in the range $0.1\lsim q\leq 1$.  Studies of the demography
and assembly of SMBHs in hierarchical cosmologies, based on the merger
histories of dark matter haloes, find broad distributions between
$10^{-2}\lsim q < 1$, generally peaking in the range $q\sim 0.1-1$
\citep[e.g.][]{Volonteri+2003,Sesana+2005,Lippai+2009,Sesana+2012}.
\cite{2008ApJ...672...83M}, \cite{2013MNRAS.436.2997D} and
\cite{2017MNRAS.466.1170M} have made use of 2D grid-based
hydrodynamical simulations to study SMBHBs embedded in thin $\alpha$
discs \citep{1973A&A....24..337S}, and included a discussion of the
disc torques.  However, these simulations excised the innermost region
surrounding the binary, and imposed an inner boundary condition,
neglecting potentially important gas dynamics and torque contributions
from inside the excised region. \cite{2014ApJ...783..134F} included
the innermost region in their simulated domain, but their study did
not focus on the binary-disc interaction, and did not present
measurements of the gas torques from that region. Closest to the
present paper are the previous studies by \cite{2009MNRAS.393.1423C}
and \cite{2012A&A...545A.127R}, both of which followed the interaction
between SMBHBs with $q=1/3$ and a self-gravitating circumbinary disc,
using 3D SPH simulation.  We will present a comparison to these
studies below.

In the present work, we use a 2D grid-based code to evolve a thin
(aspect ratio $h/r=0.1$) $\alpha$ disc for a few viscous times, until
the system has relaxed to a quasi-steady state. We use \texttt{DISCO},
a high-resolution, finite-volume moving-mesh code
\citep{2011ApJS..197...15D,2012ApJ...755....7D,2013ApJ...769...41D},
which allows us to not impose an inner boundary condition, and
therefore to include the BHs in the computational domain.  Our main
goal is to accurately measure the disc-binary torques, including
contributions from gravitational torques, as well as effective torques
from the accretion of mass and angular momentum by the BHs.  Accretion
is implemented with a sink prescription; we perform numerical
experiments to study how the mass removal rate in the sinks affects
the global structure of the circumbinary disc and the torques exerted
on the binary.

This paper is organized as follows. In \S~2, we summarise our
numerical methods. In \S~3, we present our main results on the disc
torques. We also determine the dependence of the disc morphology and
the disc torque on the sink time scale $\tau_{\rm sink}$.  In \S~4, we discuss
the implications of our results, list some caveats, and outline topics
for future work. Finally, in \S~5, we summarise our conclusions.
Throughout this paper, we adopt units in which Newton's constant, the
binary mass, and the binary separation are $G=M_{\rm bin}=2\times
M_{\rm bh}=a=1$, and the binary's orbital period is $t_{\rm
  bin}=2\pi$.

\section{Numerical Setup}

\subsection{Summary of Simulation Setup}

The model and numerical setup of this work closely resembles that in
our previous work \citep{2014ApJ...783..134F,2015MNRAS.446L..36F}. We
only briefly outline the key aspects and some modifications here, and
refer the reader to the above papers for further details. All
simulations are performed using the \texttt{DISCO} code in 2D; a
recent version of this code, including 3D magnetohydrodynamics
functionality, is publicly available \citep{DISCOcode,DISCOrelease}.

Our initial disc configuration is similar to that in
\cite{2015MNRAS.446L..36F}.  We modify the standard steady-state
Shakura\textendash Sunyaev disc profile by exponentially reducing the
density and pressure at small radii by a factor of
$\exp[-(r/r_{0})^{-10}]$ to create a hollow cavity within
$r_{0}\equiv2.5a$,
\begin{equation}
\Sigma(r)=\Sigma_{0}\exp[-(r/r_{0})^{-10}]r^{-0.5}.
\end{equation}
We do not include self-gravity; this assumption is justified for
compact binaries in the inner regions of the disc. As a result, the
normalization of the surface density $\Sigma_{0}$ does not affect the
simulation and is arbitrary. For convenience, we set $\Sigma_{0}=1$ in
code units.

We simulate the hydrodynamical evolution of a thin circumbinary disc
by solving the vertically averaged, 2D Navier\textendash Stokes
equations. The pressure is set to enforce a locally isothermal
equation of state of the form
\begin{equation}
c_{s}=(h/r)\left(\frac{M_{1}}{r_{1}}+\frac{M_{2}}{r_{2}}\right).
\end{equation}
Throughout this work, we choose $h/r=constant=0.1$. We employ a
standard $\alpha$-law viscosity \citep{1973A&A....24..337S} of the form
%
%\begin{equation}
$v=\alpha c_{s}^{2}\Omega_{k}^{-1}$, 
%\end{equation}
%
with $\alpha=constant=0.1$ (see Appendix~A in
\citealt{2013MNRAS.436.2997D} and \citealt{2014ApJ...783..134F} for
detailed descriptions of the viscosity implementation in polar
co\"ordinates).

In each of our simulations, we assume that the disc mass is small
compared to the binary mass, $M_{\rm disc}\ll M_{\rm bin}$ so that we
may neglect the self gravity of the gas, as well as changes in the
binary's orbital elements due to torques from the gas. In this work,
we further focus on the equal-mass binary case ($q=1$).

In order to ensure that a quasi-steady state has been established over
the spatial region of interest in our simulations, we evolve the
system for 2.5 viscous timescales at the radius of the cavity wall at
$r\approx 2a$,
\begin{equation}
  \label{eq:tvis1}
  t_{\rm vis}=\frac{2}{3}\left[\alpha\left(\frac{h}{r}\right)^{2}\Omega(r)\right]^{-1}\sim
  300\left(\frac{r}{2a}\right)^{3/2}t_{\rm bin},
\end{equation}
where $\Omega(r)$ is the Keplerian angular frequency in the disc at
the radial distance $r$ from the binary's center of mass and $t_{\rm
  bin}$ is the binary's orbital period.

\subsection{Accretion Prescription}

To mimic accretion onto the individual BHs, we follow the sink
prescription used in \cite{2014ApJ...783..134F} and
\cite{2015MNRAS.446L..36F}, but with some refinements.  At each
time-step, the gas surface density in cells within a distance $r_{\rm
  sink}$ (measured from each individual BH) is reduced uniformly by a
fraction $\frac{dt}{\tau_{\rm sink}}$, where $dt$ is the time-step, and the sink
time-scale $\tau_{\rm sink}$ is a parameter to control the mass removal rate,
defined as
\begin{equation}
\left(\frac{d\Sigma}{dt}\right)_{\rm sink}\equiv -\frac{\Sigma}{\tau_{\rm sink}}.
\end{equation}

In this work, we use $r_{\rm sink}=0.1a$ as the fiducial value. In
our previous work \citep{2014ApJ...783..134F}, $\tau_{\rm sink}$ was set at
the local viscous time based on the $\alpha$ viscosity law, applied
to the individual minidiscs,
\begin{equation}\label{eq:tvisc}
  t_{\rm vis}=\frac{2}{3}\frac{r^{2}}{\nu}=\frac{1}{3\pi}\frac{1}{\alpha(h/r)^{2}}
  \left(\frac{r}{a}\right)^{3/2}u^{-1/2}t_{\rm bin}.
\end{equation}
Here $r$ is the distance to each individual accreting BH (analogous to
Eq.~\ref{eq:tvis1} for the circumbinary disc), $u=1/2$ is the ratio of
the BH mass to the total binary mass, and $t_{\rm bin}$ is the binary
orbital period (which is $2\pi$ in code units).

In this work we choose $\tau_{\rm sink}$ to be a constant inside the sink radius,
and perform a suite of simulations with a range of values of $\tau_{\rm sink}$.
This allows us to examine the relationship between the binary-disc
torques and the mass removal rate in the sink.  When we model the
minidisc as a thin $\alpha$ disc, we adopt $\tau_{\rm sink}=t_{\rm vis}(r_{\rm
  sink})$ as the sink time-scale.

\subsection{Measuring Torques}

In our simulations, the mass and velocity of the black hole binary are
both fixed. In other words, we calculate the torque exerted on the BHs
by the circumbinary disc, but do not include the impact of these
torques on the binary's orbital elements.  This is justified as long
as the torques are small enough so that the binay's orbit does not
evolve during the simulation.

Similar to \cite{2012A&A...545A.127R} and \cite{DK2017},
the total torque on the binary is computed by tracking both the
gravitational and accretion torques,
\begin{equation}
T_{\rm tot}=T_{\rm gr}+T_{\rm acc}.
\end{equation}
The gravitational torque is given by summing the gravitational force
$f_{\rm gr}$ exerted by the fluid elements in each computational cell
on the individual BHs,
\begin{equation}
T_{\rm gr}=\sum_{\rm cells} r_{\rm bh}\times f_{\rm gr}.
\end{equation}
For the accretion torque, we measure the linear momentum of the gas
removed in each simulation cell inside the sink, relative to the
accreting BHs,
\begin{equation}
T_{\rm acc}=\sum_{\rm sink} r_{\rm bh}\times\dot{m_{i}}(v_{i}-v_{\rm bh}),
\end{equation}
where $v_i$ and $v_{\rm bh}$ are the velocities of the gas and of an
individual BH (as measured in the inertial frame).
With the above definitions, $T_{\rm tot}$ denotes the time derivative
of the specific angular momentum $\ell$ of the binary,
\begin{equation}
M_{\rm bin}\times\frac{d\ell}{dt}=T_{\rm tot},
\end{equation}
where $M_{\rm bin}=M_1+M_2$ is the total binary mass ($M_{\rm bin}\equiv 1$
in code units, as stated above).

\section{Results}

\begin{figure}
\includegraphics[scale=0.35]{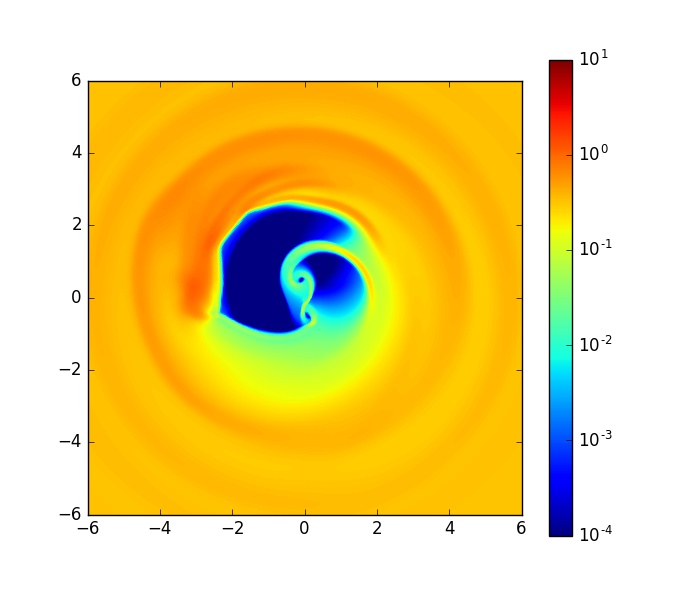}
\includegraphics[scale=0.35]{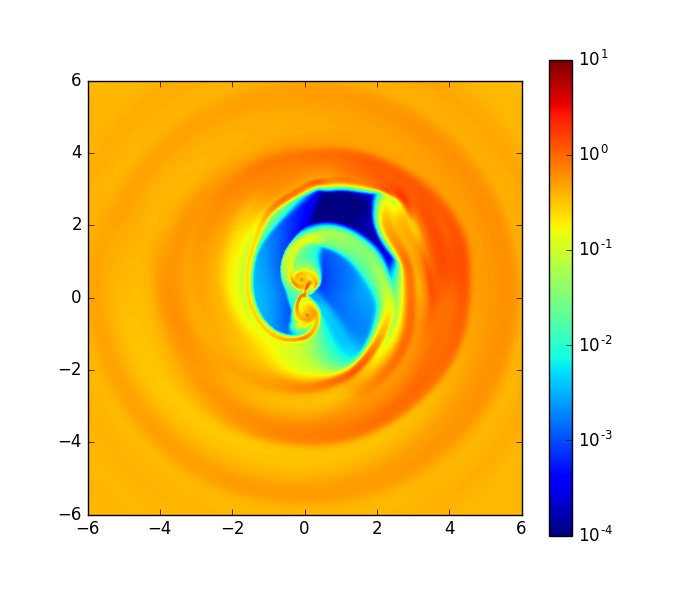}
\protect\caption{Snapshots of the surface density $\Sigma$ at
  $t\approx 800 t_{\rm bin}$.  The contours show $\Sigma$ in units of
  the initial $\Sigma_0$, on a logarithmic scale indicated by the
  side-bars. The x/y axes are in units of binary separation. The
  orbital motion of both the binary and the disc are in the
  counterclockwise direction. The upper panel is for a rapid sink
  (with mass-removal time scale $\tau_{\rm sink}=0.0125$ or
  0.002$t_{\rm bin}$), whereas the lower panel is for a slower sink
  ($\tau_{\rm sink}=5.0$ or 0.64$t_{\rm bin}$). $t_{\rm bin}=2\pi$ in
  code units.}
\label{fig:density_snap}
\end{figure}

Snapshots of the surface density from simulations with fast
($\tau_{\rm sink}=0.0125$, or 0.002$t_{\rm bin}$) and slow ($\tau_{\rm sink}=5.0$, or
0.64$t_{\rm bin}$) mass removal rate are shown in
Figure~\ref{fig:density_snap}. As discussed below, the former sink is
unphysically fast and represents a numerical experiment; the latter
value is chosen to correspond to $\alpha=0.1$.  In both cases, we
confirm the findings in many previous studies (beginning with
\citealt{al94,al96} and with recent works including, e.g.
\citealt{2008ApJ...672...83M}, \citealt{2009MNRAS.393.1423C},
\citealt{2013MNRAS.436.2997D}, \citealt{0004-637X-749-2-118},
\citealt{2012ApJ...755...51N}, \citealt{2012A&A...545A.127R}, and
\citealt{2014ApJ...783..134F}), namely that a low-density, lopsided
cavity forms around the binary, while narrow accretion streams connect
the cavity wall to the minidiscs around the individual BHs. Most of
the material in the narrow streams does not accrete onto a BH, but
rather is flung outward. The outward going streams create an overdense
lump at the cavity wall.  As the figure shows, the more rapid sink
(top panel) depletes the gas near the BHs, as well as in the entire
inner region of the circumbinary disc, compared to the slower sink
(bottom panel).

\begin{figure}
\includegraphics[scale=0.5]{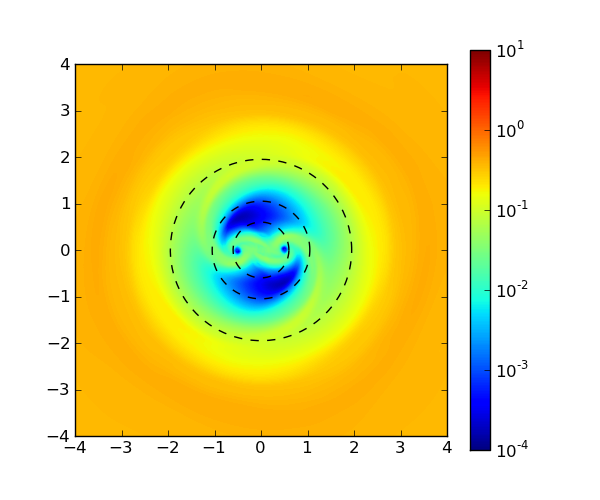}
\includegraphics[scale=0.5]{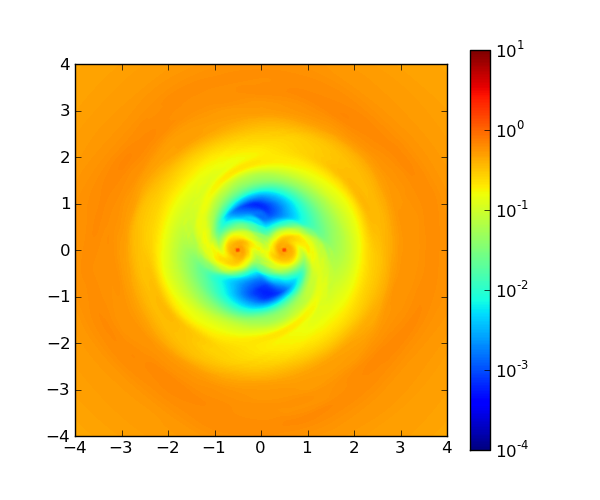}
\protect\caption{Time-averaged surface density
  profiles. Time-averaging is taken over \textasciitilde{}100 orbits,
  and in a co-rotating frame where the BHs are at the fixed
  coordinates (-0.5,0) and (+0.5,0).  As in
  Fig.~\ref{fig:density_snap}, the upper [lower] panel is for a rapid
  [slower] sink, with a mass-removal time scale of $\tau_{\rm
    sink}=0.0125$ [$5.0$]. The three dashed circles have radii of 0.6,
  1.05 and 1.95$a$, and demarkate annular regions in which the net
  gravitational torques switch sign from positive (0.6$<r<$1.05) to
  negative (1.05$<r<$1.95).}
\label{fig:average_density_snap}
\end{figure}

In Figure~\ref{fig:average_density_snap} we show the time-averaged
surface density profiles in a frame co-rotating with the binary. We
find again that with a faster sink, the surface density of the
minidiscs, of the accretion streams and of the circumbinary discs near
the cavity wall, are all visibly reduced. The three dashed circles in
the top panel have radii of 0.6, 1.05 and 1.95$a$. In the annulus
between the latter two circles, the azimuthal phase of the accretion
stream trails behind the binary.  As a result, the gravitational force
between the binary and these portions of the accretion streams exert a
negative torque on the binary (as further illustrated in
Figure~\ref{fig:surface_torque_density} below). By comparison, in the
annulus between the first two circles, the tails of the narrow streams
are ahead of the individual BHs, and therefore exert a positive
gravitational torque on the binary. Comparing the shape of the
minidiscs in the top and bottom panels, we find that the minidiscs in
the fast-sink case are less dense overall; they are furthermore less
round and symmetric than in the slow-sink case.  Finally, an important
conclusion from this figure is that the fast sink preferentially
depletes the gas density behind (versus ahead of) the individual BHs
(i.e. in the annulus $0.6<r<1.05$).

\begin{figure}
\includegraphics[scale=0.3]{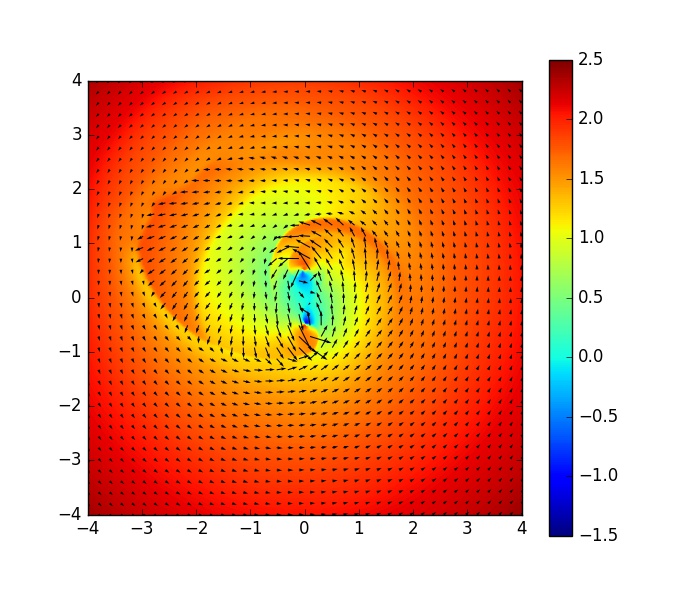}
\includegraphics[scale=0.3]{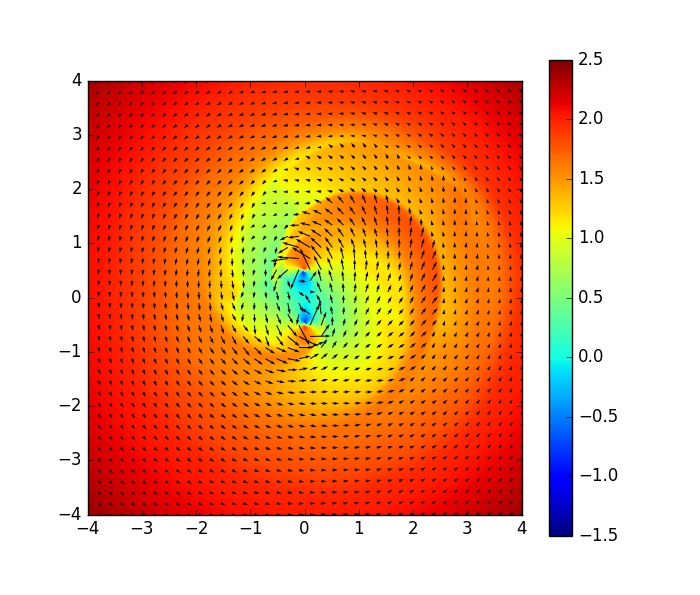}
\protect\caption{Snapshots of the distribution of specific angular
  momentum $|\vec{r}\times \vec{v}|$. The snapshots are taken from the
  same runs and at the same time as in Figure~\ref{fig:density_snap},
  and as in the previous figures, the top [bottom] panel is for a fast
  [slow] sink. Black arrows denote the local velocity field for the
  disc gas. The prominent streams travel outward, and carry angular
  momentum from the binary to the circumbinary disc.}
\label{fig:specific_angular_momentum}
\end{figure}

As discussed in several previous works \citep{2008ApJ...672...83M,
  2012A&A...545A.127R, 0004-637X-749-2-118, 2013MNRAS.436.2997D,
  2014ApJ...783..134F}, most of the material in the narrow streams
gains angular momentum from the faster moving BHs and is flung back
towards the cavity wall.  In
Figure~\ref{fig:specific_angular_momentum}, we plot contours of
specific angular momentum in runs with both fast and slow sinks (top
and bottom panel, respectively). These confirm that the specific
angular momentum of the material forming the narrow streams is larger
than that of the cavity wall, and also that overall, the streams have
a net outward-going velocity.  This suggest that the narrow streams
could extract angular momentum from the binary and deposit it into the
circumbinary disc, {\em effectively acting as media to carry angular
  momentum from the binary to the gas disc.}

\begin{figure}
\includegraphics[scale=0.5]{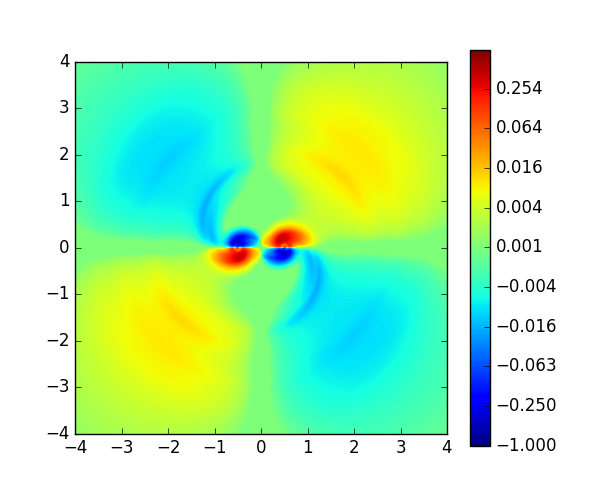}
\includegraphics[scale=0.5]{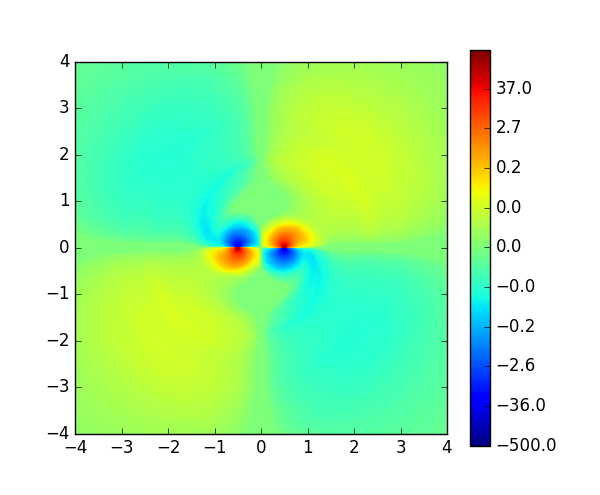}
\protect\caption{Time-averaged
  distributions of the gravitational torque surface density, exerted
  by the gas disc on the binary. The sink time scales in the top and
  bottom panels are $\tau_{\rm sink}=0.0125$ and 5.0, as in previous figures. The
  net torque is positive in the top panel and negative in the bottom
  panel.  Note the different scales in the two panels.}
\label{fig:surface_torque_density}
\end{figure}

\begin{figure}
\includegraphics[scale=0.45]{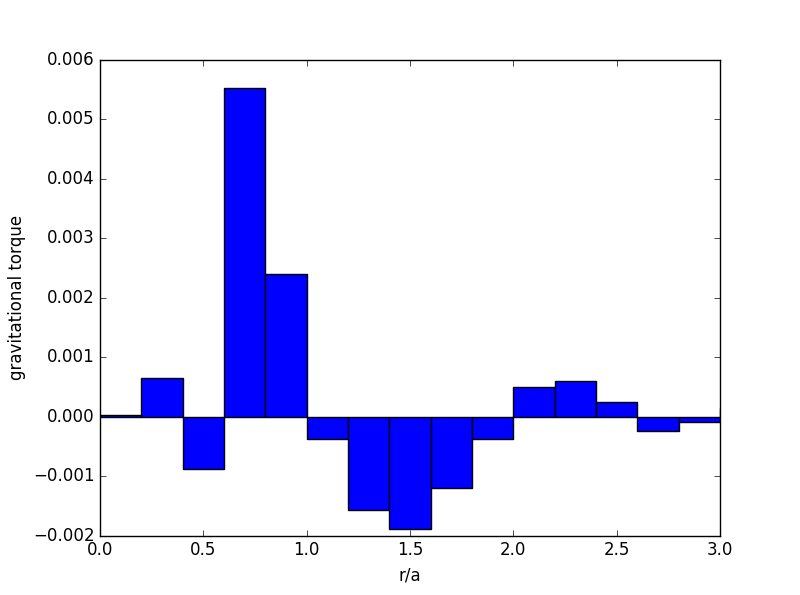}
\includegraphics[scale=0.45]{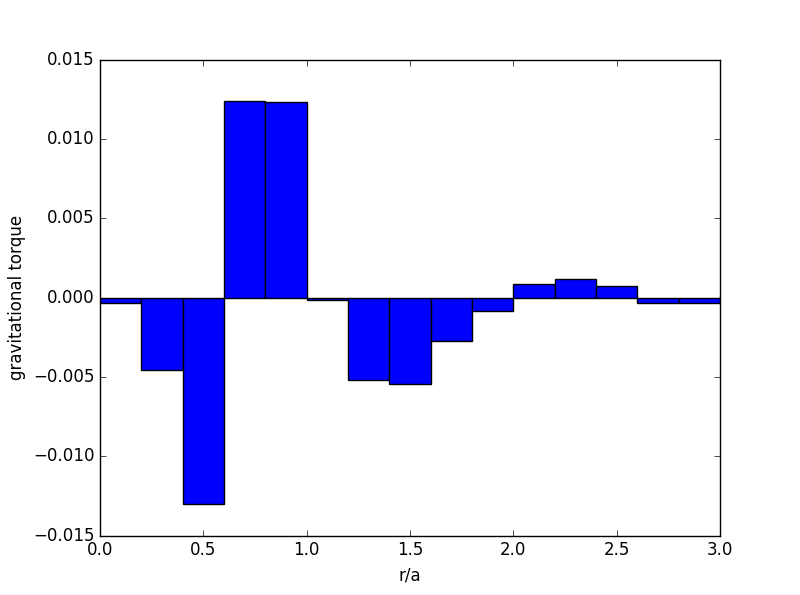}
\protect\caption{Radial distribution of the azimuthally-averaged
  gravitational torque density, corresponding to the runs in Fig.~\ref{fig:surface_torque_density}.
% $\tau_{\rm sink}=0.0125$ (effective $\alpha=238.51$) and $\tau_{\rm sink}=5.0$(effective
%  $\alpha=0.59$) . Binary period $t_{\rm bin}=2\pi$ in the code unit.
  The contribution from the region $r>3a$ is negligible in all of the
  runs. The net gravitational torques are 0.3$\dot{M}$ and
  -0.68$\dot{M}$ in the top and bottom panels, respectively.}
\label{fig:Radial_distribution_gravitational_torque}
\end{figure}

In Figure~\ref{fig:surface_torque_density}, we plot the 2D
distribution of the gravitational torque surface density, and in
Figure~\ref{fig:Radial_distribution_gravitational_torque}, we plot the
radial distribution of the azimuthally-averaged gravitational torque
density. The previous works by \cite{2008ApJ...672...83M} and
\cite{2014ApJ...783..134F} calculated the torques only in the region
$r>1.0a$. Our results are qualitatively similar to these studies in
this region. In particular, we find significant negative torques from
the disc material in the annulus between $a<r<2a$, corresponding to
the extended faint blue streams in
Figure~\ref{fig:surface_torque_density}. The total torque from the
outer regions $r>a$ is also negative, with $r>3a$ contributing
negligibly to this total. However, we find a significant positive
torque in the annulus $0.6a<r<a$.
Figure~\ref{fig:surface_torque_density} further quantifies the point
mentioned above, namely that this positive torque can be attributed to
the compact tails of the narrow streams. These can be seen as the
prominent red clumps in the figure, located near and ahead of the
individual BHs. In the fast sink case, the contribution from this
region dominates, making the total torque positive. In the slow sink
case, the negative torques from the more massive minidiscs roughly
cancel the positive torques from the tails of the narrow streams, and
the result is a net negative torque.

\begin{figure}
\includegraphics[scale=0.35]{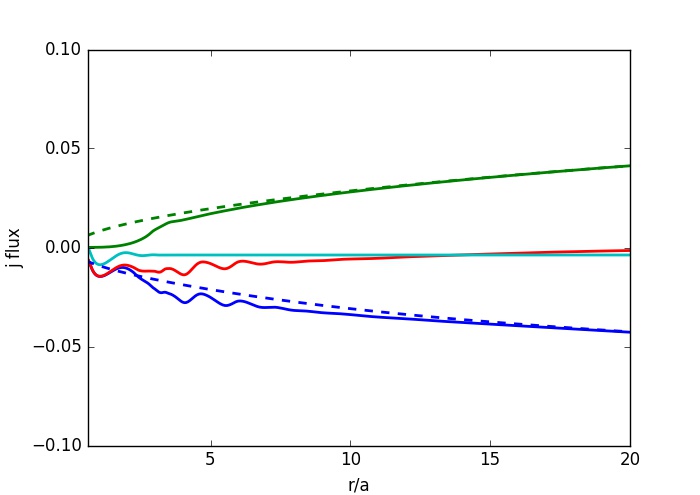}
\includegraphics[scale=0.35]{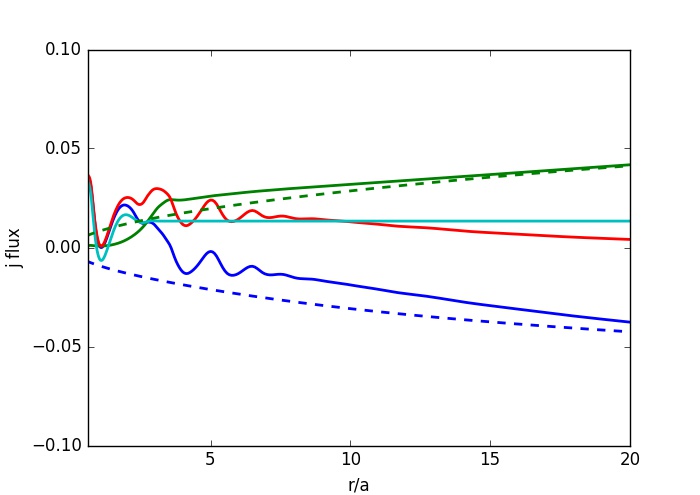}
\protect\caption{Radial profiles of angular momentum flux for a fast
  (top; $\tau_{\rm sink}=0.0125$) and slow sink (bottom panel;
  corresponding to $\alpha=0.1$).  The fluxes corresponding to
  viscosity, gravitational torques, and advection are shown by the
  green, cyan and blue curves, respectively.  The red curve is the sum
  of the green and blue curves. The dashes curves show the viscous and
  advected fluxes in a reference single-BH simulation with the same BH
  mass and disc parameters. In the bottom panel, the binary is loosing
  angular momentum to the disc, with this excess angular momentum
  transported outward by enhanced viscosity (at $r\gsim 3a$).  In the
  top panel, the binary is gaining angular momentum from the disc,
  with reduced viscosity allowing net inward angular momentum
  transfer.}
\label{fig:angular_momentum_flux}
\end{figure}

In Figure~\ref{fig:angular_momentum_flux}, we plot the radial angular
momentum flux profiles for both fast and slow sinks.  We show,
separately the fluxes corresponding to advection (blue), and to the
viscous (green) and gravitational torques (cyan). The accretion
torques are not shown explicitly, but correspond to a source term near
the BHs (they are negligible in the lower panel, but account for the
visible constant offset between the red and cyan curves at $r\approx
0$ in the upper panel).  With the fast sink, the net torque exerted on
the binary is positive, implying that angular momentum is transported
from the disc to the binary. With the slow sink, the situation is the
reverse, and angular momentum is transported outward from the binary
to the disc.

We find that inside the cavity, the viscous angular momentum flux is
close to zero, and advection is the only effective way to transfer
angular momentum.  With a slow sink, the advected angular momentum
flux turns positive in the cavity. This implies that the narrow
streams in the cavity are able to carry angular momentum outward to
the cavity wall.  When compared to the analogous results from a
single-BH reference disc (shown by the dashed curves), we find that
both the viscous and advected fluxes shift up or down based on the
direction of the net angular momentum flux. However, at large radii,
the viscous flux converges to the single-BH solution faster than the
advected flux.

Overall, we conclude that when the net torque on the binary is
negative (bottom panel), the angular momentum of the binary ends up
being deposited into the circumbinary disc, and transported outward
first by gravitational torques and advection (in the inner regions;
$r\lsim 3a$) and then by the elevated viscosity (farther out; $r\gsim
3a$).  For the fast sink (top panel), the behavior is the opposite,
with reduced viscosity enhancing the inward advection of angular
momentum, which ultimately results in a net transfer of angular
monentum from the disc onto the binary.

\begin{figure}
\includegraphics[scale=0.3]{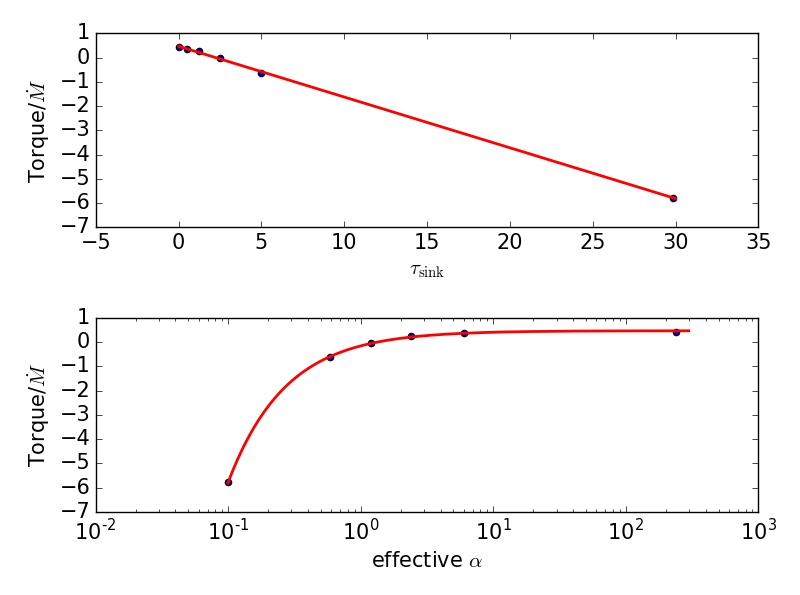}
\protect\caption{{\em Top panel}: total specific torque on the binary,
  in units of $\dot{M}$, as a function of the sink time-scale
  $\tau_{\rm sink}$.  The red line is a best-fit to the simulation
  results (black points), and has a slope and intercept of -0.209 and
  0.437, respectively.  {\em Bottom panel:} total torque as a function
  of the effective viscosity parameter $\alpha$ (calculated from
  Eq.~\ref{eq:tvisc}).}
\label{fig:torque_vs_tau}

\end{figure}

In Figure~\ref{fig:torque_vs_tau}, we show the net specific torque on
the binary, in units of $\dot{M}$, as a function of the sink time
scale $\tau_{\rm sink}$.  We find a clear linear relationship with longer sink
timescales yielding more negative specific torque.  The total torque
exerted on the binary is well approximated by the equation:
\begin{equation}
T_{\rm tot}=(-0.209\tau_{\rm sink}+0.437)\dot{M}.
\end{equation}
When $\tau_{\rm sink}$ is below a critical value of $\tau_{\rm sink,crit}=2.1$, the
binary feels a net positive torque from the disc, and is driven apart
by the gas.\footnote{We here ignore the negative torques from
  gravitational waves, which could overwhelm the positive gas torque
  for sufficiently compact binaries.}  In the bottom pannel of
Figure~\ref{fig:torque_vs_tau}, we show the total torque as a function
of the corresponding effective $\alpha$ (using Eq.~\eqref{eq:tvisc} to
convert $\tau_{\rm sink}=t_{\rm visc}$ to $\alpha$).  The effective viscosity
parameter corresponding to $\tau_{\rm sink,crit}=2.1$ is $\alpha_{\rm
  crit}=1.42$.  We conclude that for sink time-scales corresponding to
physically realistic choices of viscosity, binaries are always driven
towards merger, rather than pushed apart, by the presence of the
circumbinary gas disc.

\cite{Kocsis+2012a,Kocsis+2012b}, \cite{Rafikov2012} and recently
\cite{2016ApJ...827..111R} have used simplified 1D models to study how
the surface density profile of a steady accretion disc responds when
angular momentum is injected to (or extracted from) an $\alpha$ disc
at some small radius by a binary companion (see also
\citealt{1995MNRAS.277..758S} for related earlier work).  These
analytic 1D steady-state solutions resemble standard Shakura-Sunyaev
discs, but the disc adjusts itself to transfer angular momentum
outward (or inward) at the same rate as the injection (or extraction)
rate imposed at the bottom (i.e. at a small radius, comparable to the
binary separation).  This modified angular momentum transfer rates
requires a surface density 'pile-up' (or depletion) in the inner disc
regions.

The form of this modified disc profile can be approximated by
combining equations (3) and (11) of \cite{2016ApJ...827..111R} (see
also his Figure 2). The modified surface density follows the form
\begin{equation}
\Sigma(r)=A/\sqrt{r}-B/r,
\end{equation}
where $A$ and $B$ are constants, related to one another as
\begin{equation}
B/A=\dot{J}/\dot{M}.
\label{eq:B/A}
\end{equation}
Here $\dot{J}$ is the angular momentum flux absorbed by the binary
(which can be either negative or positive), and $\dot{M}$ is the mass
accretion rate which is always positive.

\begin{figure}
\includegraphics[scale=0.45]{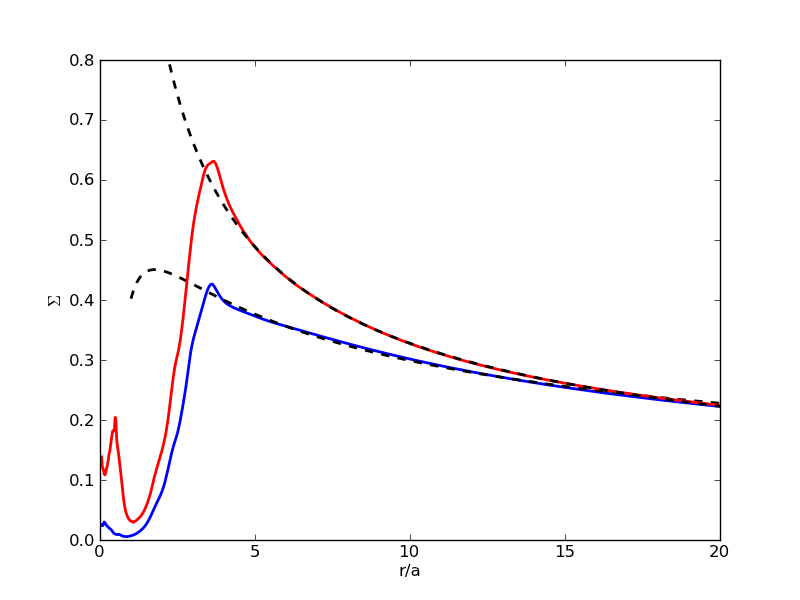}
\includegraphics[scale=0.32]{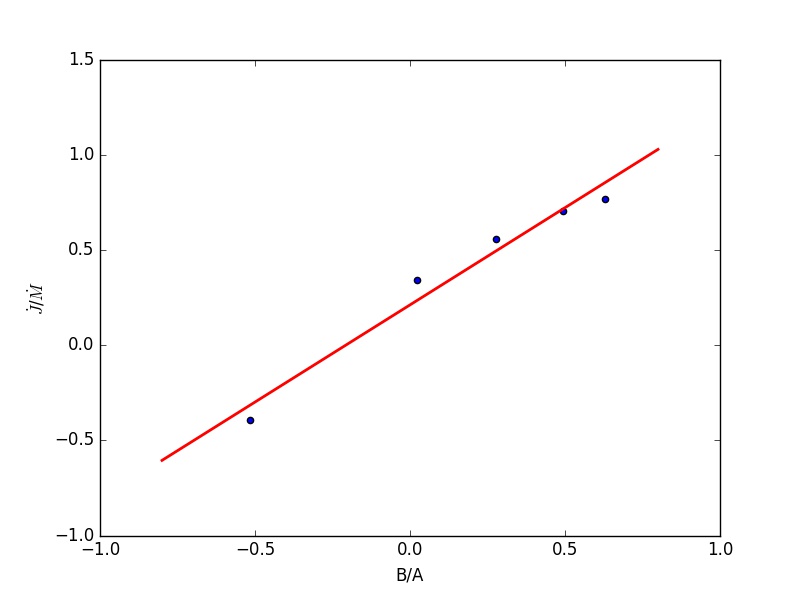}
\protect\caption{Time-averaged surface density profiles and best-fit
  analytic curves.  The red and blue curves in the top panel
  correspond to the fast and slow sinks, respectively. The black
  dashed curves follow $\frac{A}{\sqrt{r}}-\frac{B}{r}$, where $A$ and
  $B$ are free parameters fit to the simulated profiles. The bottom
  panel shows the relationship between $\langle\dot{J}/\dot{M}\rangle$
  and the ratio $B/A$ of the best-fit values. The straight red line is
  a least-squares fit to the simulation results (shown by black
  points), and has a slope of $\sim1.02$ and intercept
  $\sim0.21$. These are in good agreement with the predictions of 1D
  steady-state models for how the disc responds to binary torques (see
  text for discussion).}
\label{fig:B_A}
\end{figure}

In Figure~\ref{fig:B_A}, we show the time- and azimuthally-averaged
surface density profiles from our simulations, for both a fast (top
panel) and a slow sink (bottom panel). We find that outside the cavity
wall, $\Sigma(r)$ accurately follows the form given in
Eq.~\eqref{eq:B/A}.  In Figure~\ref{fig:B_A}, we further test the 1D
predictions, and plot the empirical relationship between
$\langle\dot{J}\rangle/\langle\dot{M}\rangle$ and the $B/A$ value
obtained from fitting the profiles found in the simulations. We find
\begin{equation}
\frac{\langle\dot{J}\rangle}{\langle\dot{M}\rangle}=1.02\left(\frac{B}{A}\right)+0.21,
\end{equation}
which matches the prediction (eq.~\ref{eq:B/A}) quite well.
Nevertheless, we caution that in order to reach a steady-state
throughout the disc, we would need to run the simulations for several
viscous times measured at the outermost disc radius. Thus,
Figure~\ref{fig:B_A} should be interpreted only as evidence for a
``temporary'' steady state in the inner regions, and for the sink time
scale affecting the global surface density structure as expected,
rather than a prediction for the long-term steady-state surface
density profile, which would be established on a much longer
time-scale.

\cite{RyanMacFadyen2017} have performed a general relativistic local
simulation of an isolated minidisc, subject to the tidal field of a
companion. Their simulation does not include gas viscosity, but shows
gravitational torques arising from non-axisymmetric structures (spiral
waves) inside the minidisc, induced by the tidal torques of the
companion BH.  They find that the minidisc has an effective
``gravitational Shakura-Sunyaev $\alpha$'' on the order of a few
$\times 10^{-2}$.  Since this effect does not significantly enhance
the effective $\alpha$ parameter of the minidisc, we proceed to assume
the physical sink time scale $\tau_{\rm sink}$ equals the viscous time
at $r_{\rm sink}$.

\begin{figure}
\includegraphics[scale=0.32]{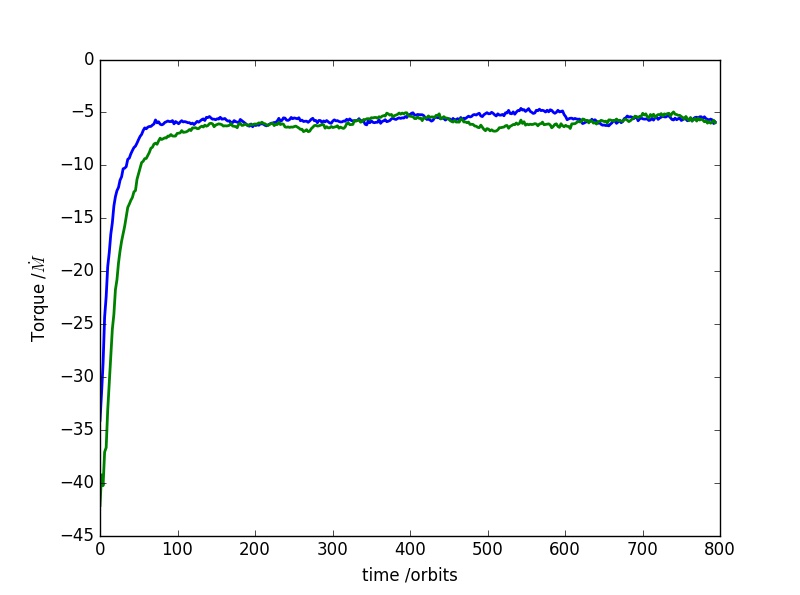}
\protect\caption{Moving time-averaged total torque {\it vs} time with
  two different choices of sink radius. The blue curve is for $r_{\rm
    sink}=0.05a$ and the green curve is for $r_{sink}=0.1a$. The time
  averaging window is 60 orbits.  The torques converge rapidly to a
  value that is insensitive to the choice of $r_{\rm sink}$.  }
\label{fig:rsink}
\end{figure}

To investigate whether the sink radius $r_{\rm sink}$ affects our
results, we have performed simulations with $r_{\rm sink}=0.1a$ and
$0.05a$. In each case, we have set the sink time scale to $\tau_{\rm sink}=t_{\rm
  vis}(r_{\rm sink})$. Figure~\ref{fig:rsink} shows the specific
torque, in units of $\dot{M}$, as a function of time in these two
simulations. The figure shows that the torques converge rapidly to a
value that is insensitive to the choice of $r_{\rm sink}$.  In both
runs, we find that the gravitational torque dominates over the
accretion torque, with $|T_{\rm acc}|/|T_{\rm tot}|\approx
0.01$.

In summary, in Table~1 we list the simulation parameters and the gas
disc torques exerted on the binary in each of our simulations.

\begin{table}
  \begin{center}
\begin{tabular}{|c|c|c|c|}
\hline 
$\tau_{\rm sink}$ & effective $\alpha$ & $r_{\rm sink}$ & $T_{\rm tot}/\dot{M}$\tabularnewline
\hline 
\hline 
0.0125 & 238.51 & 0.1 & 0.433\tabularnewline
\hline 
0.5 & 5.96 & 0.1 & 0.373\tabularnewline
\hline 
1.25 & 2.38 & 0.1 & 0.269\tabularnewline
\hline 
2.5 & 1.19 & 0.1 & -0.0271\tabularnewline
\hline 
5.0 & 0.59 & 0.1 & -0.625\tabularnewline
\hline
\hline 
29.84 & 0.1 & 0.1 & -5.77\tabularnewline
\hline 
9.621 & 0.1 & 0.05 & -5.52\tabularnewline
\hline 
\end{tabular}
\protect\caption{Summary of our simulation runs. The columns, from
  left to right, show the (i) sink time-scale, (ii) the corresponding
  effective viscosity parameter, (iii) sink radius, and (iv) total
  dimensionless specific torque.}
\end{center}
  \end{table}

\section{Discussion}

\subsection{Implications for the SMBH binary population}

In our simulations the binary orbital period is $2\pi$, and the
accretion rate $\dot{M}$ is arbitrary, depending linearly on the
surface density normalization $\Sigma_{0}$. By scaling the simulation
to $t_{\rm bin}=1$yr, and $\dot{M}=0.3\dot{M}_{\rm Edd}$ (with
$\dot{M}_{\rm Edd}\equiv L_{\rm Edd}/\epsilon c^2$, $\epsilon=0.1$
i.e. the accretion rate corresponding to the Eddington luminosity), we
obtain the binary's residence time $t_{\rm res}\equiv a/[da/dt]$ in
physical units.  We find that this time-scale due to the gas torque is
$t_{\rm res}= 3.26\times10^{6}$yr.  In our code units ($M_{\rm
  bin}=a=1$, $t_{\rm bin}=2\pi$) the conversion constant for the
accretion rate is $t_{\rm bin}/(2\pi M_{\rm bin})$. Therefore if
$\dot{M}/\dot{M}_{\rm Edd}$ is fixed, in our code units $\dot{M}$ is
proportional to $t_{\rm bin}$ and independent of $M_{\rm bin}$. In
code units the residence time $\widetilde{t}_{\rm res}\varpropto
T_{tot}^{-1}\varpropto\Sigma_{0}^{-1}\varpropto\dot{M}^{-1}\varpropto
t_{\rm bin}^{-1}$ and the physical residence time $t_{\rm
  res}=\widetilde{t}_{\rm res}\times t_{\rm bin}/2\pi$. Thus the
inferred residence time in physical units does not depend on $M_{\rm
  bin}$ or $t_{\rm bin}$.  The same conclusion would follow by noting
that the residence time in physical units would scale as $t_{\rm
  res}\propto$ (angular momentum/torque) $\propto a/(\Sigma_0 t_{\rm
  bin})$, with $a$, $\Sigma_0$, and $t_{\rm bin}$ in physical units.
However, the condition $\dot{M}=const\times\dot{M}_{\rm Edd}$ implies
$\Sigma_0\propto a/t_{\rm bin}$, making $t_{\rm res}$ a constant.
This counter-intuitive conclusion is a result of our scale-free
simulation set-up, and would no longer hold in the presence of
realistic cooling/heating or other physics introducing a physical
scale.

Further choosing a total binary mass $M_{\rm bin}=10^{6-9}~\msun$, the
inspiral time-scale due to gravitational wave emission is $t_{\rm
  GW}=5.2\times10^{8-3}$yr \citep{Peters1964}.  We conclude that the
scaled gas-driven inspiral time, implied by our simulations, is
comparable to (or shorter than) the GW-driven inspiral time for
binaries with total mass $M_{\rm bin} \lsim 10^7~{\rm M_\odot}$.

Pulsar timing arrays are sensitive to $\sim$nHz-frequency
gravitational waves generated by SMBHBs with orbital periods between
$\sim$0.1-10 yr. Because gas torques can promote orbital decay of the
binary much more efficiently than pure GW emission, the expected
number of binaries, and the GW background (GWB) at these separations
should be reduced accordingly \citep{2011MNRAS.411.1467K,Kelley+2017}.  Although
we find that for the $10^{8-9}~\msun$ binaries which dominate the GWB,
the gas torques are subdominant, we expect the impact of the gas
torques to increase for smaller mass ratios (because gas torques
increase, whereas GW torques decrease, for smaller $q$;
e.g. \citealt{2009ApJ...700.1952H}).  \cite{1995MNRAS.277..758S} (see
also \citealt{Kocsis+2012a,Kocsis+2012b}) have proposed analytical
models analogous to Type-II migration, in which migration rates are
tied to the disc viscous time, but are slowed down by a factor related
to the ratio $M_{\rm 2}/M_{\rm disc}$, when this ratio is above unity.

For a steady-state $\alpha$ disc with $M_{\rm
  bin}=2\times10^{6-9}M_{\odot}$, and accretion rate
$\dot{M}=0.3\dot{M}_{\rm Edd}$, the viscous time $t_{\rm vis}$ at
$r=0.5a$ is approximately $(8\times10^{4}$-$10^{6}$) yr, and
the predicted residence time $t_{\rm res}$ is approximately
(50-3)$t_{\rm vis}\approx (4-3)\times 10^{6}$ years (see
\citealt{2009ApJ...700.1952H} for a detailed calculation). This turns
out to be very close to the result $t_{\rm res}\sim 3\times 10^{6}$ years
we infer here by scaling the torques measured in our simulations.  We
emphasize that this good agreement appears to be a coincidence,
because in the 1D models, the torques are assumed to arise from the
overdense cavity wall of the circumbinary disk, whereas in our
simulations, the torques are dominated by the gas near the individual
BHs.

We note further that a steady-state $\alpha$ disc with an accretion
rate of $0.3\dot{M}_{\rm Eedd}$ has a scale-height of
$H/r\sim10^{-3}$, and is much thinner than the disc in our simulations
$(H/r=0.1)$. While simulating discs numerically with $H/r\sim10^{-3}$
is currently prohibitively challenging, our choice of $H/r=0.1$ is
typical for the values employed in previous works
\citep{al94,2008ApJ...672...83M,2013MNRAS.436.2997D,2014ApJ...783..134F}
and allows for a direct comparison with those works.  We expect that
the dimensionless torques measured in the simulations will depend on
$H/r$, and we intend to investigate this dependence in future work.

We have found that the sink prescription and time-scale affect the
magnitude (and even the sign) of the torques on the binary. Slower
sinks result in more negative torques on the binary, while faster
sinks result in diminished negative (or even positive) torques. During
the late stages of the merger, the Schwarzschild radius $r_{S}$ is not
much smaller than the binary seperation. A more physically realistic
sink prescription during this stage would be to adopt the dynamical
time as the sink time-scale $\tau_{\rm sink}$, and the innermost
stable circular orbit ($r_{\rm ISCO}$) as the sink radius $r_{\rm
  sink}$.  However a more careful treatment of relativistic effects,
e.g. through the use of the Paczynski-Wiita potential
\citep{1980A&A....88...23P} is required in this case.

\subsection{Comparison with previous works}

The nature of the torques and the corresponding binary evolution,
arising from the disc-binary interaction, has been previously
investigated by \cite{2009MNRAS.393.1423C}, \cite{2012A&A...545A.127R}
and \cite{2017MNRAS.466.1170M} as mentioned in the Introduction.
\cite{2009MNRAS.393.1423C} and \cite{2012A&A...545A.127R} have both
simulated self-gravitating gaseous discs of mass $M_{\rm
  disc}=0.2M_{\rm bin}$ and BH mass ratio $q=1/3$. The disc is Toomre
unstable, and self-gravity acts as a source of viscosity that can
transport angular momentum outwards. In their SPH simulations, gas
particles in the sink are absorbed by the black hole immediately,
which corresponds to a sink time scale $\tau_{\rm sink}=0$. Similar to
our results, \cite{2012A&A...545A.127R} find that the gravitational
torque changes sign at $r=a$ (positive at $r<a$ and negative at
$r>a$). They also find that the accretion torque alone would shrink
the binary seperation significantly, while in our work the accretion
torque drives the binary apart, but is generally very weak, and
unimportant compared to the gravitational torque (a $\lsim 1\% $
correction). Furthermore, in \cite{2012A&A...545A.127R}, the binary is
driven together by the disc, and the residence time is approximately
8000 binary orbits (taking their run closest to ours, i.e. ``iso10''
in their Fig. 3).

These differences are partly explained by the fact that in
\cite{2012A&A...545A.127R}, the typical accretion rate is much higher
than the Eddington rate (20-40 $\dot{M}_{\rm Edd}$ for the secondary
and 4-7 for the primary when $M_{\rm primary}=2\times 10^6\msun$ and
$a=0.039$pc). We have found that the total torque exerted on the
binary sensitively depends on the sink timescale, and with a fast sink
the binary is driven apart by the gas. If we re-scale our simulations
to the same $\dot{M}$, $M_{\rm bin}$, and seperation $a$, and we adopt
$\tau_{\rm sink}=t_{visc}$, the residence time we obtain is
approximately 250 binary orbits, which is around 30 times
smaller. This difference may come from the different binary mass
ratio, equation of state, disc thickness, as well as from the effect
of self-gravity.  Moreover, we suspect that for a more physical
(i.e. less rapid) sink prescription was employed, the migration rate
$-\frac{da}{dt}$ in \cite{2009MNRAS.393.1423C} and
\cite{2012A&A...545A.127R} would be faster, and closer to our value.

Finally, we note that \cite{2017MNRAS.466.1170M} perform a series of
2D viscous hydrodynamics simulations of circumbinary discs, but with
the inner regions ($r<a$) excised from the simulation domain. They
conclude that the binary feels a net positive torque.  This positive
torque arises from accretion, and is a direct result of the assumption
that the specific angular momentum of the dic gas near $r\approx a$
ends up being accreted by the binary. As we have seen in our
simulations which resolve the inner regions, this is not the
case. Rather than being deposited into the binary, the large angular
momentum of this gas is returned to the disk, via narrow streams, in a
'gravitational slingshot' mechanism.

\subsection{Caveats and future work}

In future work, we intend to relax several assumptions, while
expanding the parameter space of our simulations. In this paper, we
have focused on a mass ratio $q=1$, viscosity $\alpha=0.1,$ and disc
aspect ratio $H/r=0.1$. In the future, we intend to to study how these
parameters affect the torques and the evolution of the binary's
orbital parameters.  In this paper we have further assumed a locally
isothermal equation of state, with the temperature set to the same
value for the circumbinary disc and for both minidiscs. This
assumption should be relaxed by incorporating a more sophisticated
treatment of the thermodynamics, so that the minidiscs are allowed to
heat or cool independently \citep[e.g.][]{2015MNRAS.446L..36F}.
Realistic circumbinary discs will also have a range of thicknesses,
and we intend to investigate the dependence of gas torque on $H/r$
\citep[e.g.][]{Ragusa+2016}.  In this paper, the binary moves on a fixed,
prescribed circular orbit. In future work, this assumption should also
be relaxed, in order to measure the growth of eccentricity due to the
binary-torque interaction \citep[e.g.][]{2009MNRAS.393.1423C} and
self-consistently couple the evolution of the binary and the disc
\citep[e.g.][]{DK2017}.  If significant eccentricity is imparted to the
binary, it may have observational consequences for GW observations,
especially for the GWB measured by PTAs. Finally, our simulations are
in 2D. We expect that the 3D vertical structure will modify the
structure of the minidisc and the accretion streams, and inevitably
have a strong affect the gas torques.

\section{Conclusions}

We have performed hydrodynamic simulations of thin, circumbinary
accretion discs using a moving-mesh, finite volume code with the BHs
present on the simulated grid.  Our discs are locally isothermal,
co-rotating in the plane of the binary, and we employ an $\alpha$
viscosity prescription.  The binaries follow a fixed, prescribed
circular orbit. We carried out a detailed study of the gas torques
exerted on the BHB. We performed a numerical experiment, to study the
effect of the mass removal rate in the sink prescription used to model
accretion onto the individual BHs, on both the gas torques and on the
global circumbinary disc structure. The most important conclusions of
this work can be summarized as follows:

(1) We find a strong reciprocal relationship between the mass-removal
time scale and the torques on the binary. The total torque is positive
when the mass-removal time scale $\tau_{\rm sink}$ is $\lesssim
2.5$. For slower sinks, gas accumulates near the BHs and generates
large negatives torques, making the net total torque exerted on the
binary negative. The relationship between $\tau_{\rm sink}$ and total
gas torque $T_{\rm tot}$ can be well approximated by a linear
regression. When we model the minidiscs as thin $\alpha$ discs, and
adopt a viscous time as the sink time-scale, the gas torque exerted on
the binary is always negative for $\alpha\leq 1$, and insensitive to
$r_{\rm sink}$.

(2) The gas torques are predominantly gravitational, and determined by
gas dynamics near individual BHs. The gas torques come from threee
regions: the narrow streams, tails of the narrow streams near the
minidiscs, and the minidiscs themselves. The tail of the narrow stream
contributes positive torque, while the other regions contribute
negative torque.  Rapid sinks make the tail of the narrow streams
dominate, and result in a net positive total torque. For slower sinks,
the negative torque from the material in the minidiscs behind the BHs
cancel the positive torques from the gas ahead of the BHs, and result
in a net negative torque. When we adopt the viscous time as the sink
time-scale, the accretion torques are at most a few percent of the
total torque.

(3) When are simulations are scaled to physical units, we find that
binaries are driven together on a time scale of $t_{\rm
  res}=a/|da/dt|\sim3\times10^{6}$ years. This time-scale would be
independent of $M_{\rm bin}$ or $t_{\rm bin}$, if the dimensionless
torques were independent of the disc aspect ratio $H/R$.  However,
this is unlikely, and the migration rates will in general depend on
$H/R$ (and then also on $M_{\rm bin}$ or $t_{\rm bin}$).  Our current
results suggest that the scaled gas-driven inspiral time is comparable
to (or shorter than) the GW-driven inspiral time for binaries with
total mass $M_{\rm bin} \lsim 10^7~{\rm M_\odot}$.  However, the above
dependencies, as well as the caveats listed in the previous section,
will have to be explored in future work, in order to accurately assess
the impact of gas torques on the binary population.

%For a $M_{bh}=10^{9}M_{\odot}$ binary with $t_{\rm
%  bin}=1year$ the residence time due to GW wave is $10^{4}$ times
%larger. Thus the expected number of binaries which generate nHz GW
%background may need to be modified accordingly. For a steady state
%$\alpha$ disc with $M_{bh}=2\times10^{6}M_{\odot}$,
%$\dot{M}=0.3\dot{M}_{edd}$ the viscous time at r=0.5a is around
%$t_{res}/40$ and our result is similar with residence time predicted
%by analytical model.

\section*{Acknowledgements}

We thank Daniel D'Orazio, Paul Duffell, and Geoff Ryan for useful
discussions. Financial support was provided from NASA through the ATP
grant NNX15AB19G (ZH). ZH also gratefully acknowledges support from a
Simons Fellowship for Theoretical Physics.

%%%%%%%%%%%%%%%%%%%%%%%%%%%%%%%%%%%%%%%%%%%%%%%%%%

%%%%%%%%%%%%%%%%%%%% REFERENCES %%%%%%%%%%%%%%%%%%

% The best way to enter references is to use BibTeX:

%\bibliographystyle{mnras}
%\bibliography{example} % if your bibtex file is called example.bib

% Alternatively you could enter them by hand, like this:
% This method is tedious and prone to error if you have lots of references
\bibliography{paper}

\begin{thebibliography}{}
\makeatletter
\relax
\def\mn@urlcharsother{\let\do\@makeother \do\$\do\&\do\#\do\^\do\_\do\%\do\~}
\def\mn@doi{\begingroup\mn@urlcharsother \@ifnextchar [ {\mn@doi@}
  {\mn@doi@[]}}
\def\mn@doi@[#1]#2{\def\@tempa{#1}\ifx\@tempa\@empty \href
  {http://dx.doi.org/#2} {doi:#2}\else \href {http://dx.doi.org/#2} {#1}\fi
  \endgroup}
\def\mn@eprint#1#2{\mn@eprint@#1:#2::\@nil}
\def\mn@eprint@arXiv#1{\href {http://arxiv.org/abs/#1} {{\tt arXiv:#1}}}
\def\mn@eprint@dblp#1{\href {http://dblp.uni-trier.de/rec/bibtex/#1.xml}
  {dblp:#1}}
\def\mn@eprint@#1:#2:#3:#4\@nil{\def\@tempa {#1}\def\@tempb {#2}\def\@tempc
  {#3}\ifx \@tempc \@empty \let \@tempc \@tempb \let \@tempb \@tempa \fi \ifx
  \@tempb \@empty \def\@tempb {arXiv}\fi \@ifundefined
  {mn@eprint@\@tempb}{\@tempb:\@tempc}{\expandafter \expandafter \csname
  mn@eprint@\@tempb\endcsname \expandafter{\@tempc}}}

\bibitem[\protect\citeauthoryear{{Artymowicz} \& {Lubow}}{{Artymowicz} \&
  {Lubow}}{1994}]{al94}
{Artymowicz} P.,  {Lubow} S.~H.,  1994, \mn@doi [\apj] {10.1086/173679}, \href
  {http://adsabs.harvard.edu/abs/1994ApJ...421..651A} {421, 651}

\bibitem[\protect\citeauthoryear{{Artymowicz} \& {Lubow}}{{Artymowicz} \&
  {Lubow}}{1996}]{al96}
{Artymowicz} P.,  {Lubow} S.~H.,  1996, \mn@doi [\apjl] {10.1086/310200}, \href
  {http://adsabs.harvard.edu/abs/1996ApJ...467L..77A} {467, L77+}

\bibitem[\protect\citeauthoryear{{Begelman}, {Blandford}  \& {Rees}}{{Begelman}
  et~al.}{1980}]{Begel:Blan:Rees:1980}
{Begelman} M.~C.,  {Blandford} R.~D.,   {Rees} M.~J.,  1980, \mn@doi [\nat]
  {10.1038/287307a0}, 287, 307

\bibitem[\protect\citeauthoryear{{Cuadra}, {Armitage}, {Alexander}  \&
  {Begelman}}{{Cuadra} et~al.}{2009}]{2009MNRAS.393.1423C}
{Cuadra} J.,  {Armitage} P.~J.,  {Alexander} R.~D.,   {Begelman} M.~C.,  2009,
  \mn@doi [\mnras] {10.1111/j.1365-2966.2008.14147.x}, \href
  {http://adsabs.harvard.edu/abs/2009MNRAS.393.1423C} {393, 1423}

\bibitem[\protect\citeauthoryear{{D'Orazio}, {Haiman}  \&
  {MacFadyen}}{{D'Orazio} et~al.}{2013}]{2013MNRAS.436.2997D}
{D'Orazio} D.~J.,  {Haiman} Z.,   {MacFadyen} A.,  2013, \mn@doi [\mnras]
  {10.1093/mnras/stt1787}, \href
  {http://adsabs.harvard.edu/abs/2013MNRAS.436.2997D} {436, 2997}

\bibitem[\protect\citeauthoryear{{Dotti}, {Sesana}  \& {Decarli}}{{Dotti}
  et~al.}{2012}]{Dotti:2012:rev}
{Dotti} M.,  {Sesana} A.,   {Decarli} R.,  2012, \mn@doi [Advances in
  Astronomy] {10.1155/2012/940568}, \href
  {http://adsabs.harvard.edu/abs/2012AdAst2012E...3D} {2012}

\bibitem[\protect\citeauthoryear{{Duffell}}{{Duffell}}{2016a}]{DISCOcode}
{Duffell} P.~C.,  2016a, {DISCO: 3-D moving-mesh magnetohydrodynamics package},
  Astrophysics Source Code Library - ASCL:1605.011 (\mn@eprint {ascl}
  {1605.011})

\bibitem[\protect\citeauthoryear{{Duffell}}{{Duffell}}{2016b}]{DISCOrelease}
{Duffell} P.~C.,  2016b, \mn@doi [\apjs] {10.3847/0067-0049/226/1/2}, \href
  {http://adsabs.harvard.edu/abs/2016ApJS..226....2D} {226, 2}

\bibitem[\protect\citeauthoryear{{Duffell} \& {MacFadyen}}{{Duffell} \&
  {MacFadyen}}{2011}]{2011ApJS..197...15D}
{Duffell} P.~C.,  {MacFadyen} A.~I.,  2011, \mn@doi [\apjs]
  {10.1088/0067-0049/197/2/15}, \href
  {http://adsabs.harvard.edu/abs/2011ApJS..197...15D} {197, 15}

\bibitem[\protect\citeauthoryear{{Duffell} \& {MacFadyen}}{{Duffell} \&
  {MacFadyen}}{2012}]{2012ApJ...755....7D}
{Duffell} P.~C.,  {MacFadyen} A.~I.,  2012, \mn@doi [\apj]
  {10.1088/0004-637X/755/1/7}, \href
  {http://adsabs.harvard.edu/abs/2012ApJ...755....7D} {755, 7}

\bibitem[\protect\citeauthoryear{{Duffell} \& {MacFadyen}}{{Duffell} \&
  {MacFadyen}}{2013}]{2013ApJ...769...41D}
{Duffell} P.~C.,  {MacFadyen} A.~I.,  2013, \mn@doi [\apj]
  {10.1088/0004-637X/769/1/41}, \href
  {http://adsabs.harvard.edu/abs/2013ApJ...769...41D} {769, 41}

\bibitem[\protect\citeauthoryear{{Duffell}, {Haiman}, {MacFadyen}, {D'Orazio}
  \& {Farris}}{{Duffell} et~al.}{2014}]{2014ApJ...792L..10D}
{Duffell} P.~C.,  {Haiman} Z.,  {MacFadyen} A.~I.,  {D'Orazio} D.~J.,
  {Farris} B.~D.,  2014, \mn@doi [\apjl] {10.1088/2041-8205/792/1/L10}, \href
  {http://adsabs.harvard.edu/abs/2014ApJ...792L..10D} {792, L10}

\bibitem[\protect\citeauthoryear{{D{\"u}rmann} \& {Kley}}{{D{\"u}rmann} \&
  {Kley}}{2017}]{DK2017}
{D{\"u}rmann} C.,  {Kley} W.,  2017, \mn@doi [\aap]
  {10.1051/0004-6361/201629074}, \href
  {http://adsabs.harvard.edu/abs/2017A%26A...598A..80D} {598, A80}

\bibitem[\protect\citeauthoryear{Escala, Larson, Coppi  \& Mardones}{Escala
  et~al.}{2005}]{Escala:2005}
Escala A.,  Larson R.~B.,  Coppi P.~S.,   Mardones D.,  2005, \mn@doi [The
  Astrophysical Journal] {10.1086/431747}, 630, 152

\bibitem[\protect\citeauthoryear{{Farris}, {Duffell}, {MacFadyen}  \&
  {Haiman}}{{Farris} et~al.}{2014}]{2014ApJ...783..134F}
{Farris} B.~D.,  {Duffell} P.,  {MacFadyen} A.~I.,   {Haiman} Z.,  2014,
  \mn@doi [\apj] {10.1088/0004-637X/783/2/134}, \href
  {http://adsabs.harvard.edu/abs/2014ApJ...783..134F} {783, 134}

\bibitem[\protect\citeauthoryear{{Farris}, {Duffell}, {MacFadyen}  \&
  {Haiman}}{{Farris} et~al.}{2015}]{2015MNRAS.446L..36F}
{Farris} B.~D.,  {Duffell} P.,  {MacFadyen} A.~I.,   {Haiman} Z.,  2015,
  \mn@doi [\mnras] {10.1093/mnrasl/slu160}, \href
  {http://adsabs.harvard.edu/abs/2015MNRAS.446L..36F} {446, L36}

\bibitem[\protect\citeauthoryear{{Goldreich} \& {Tremaine}}{{Goldreich} \&
  {Tremaine}}{1980}]{GT80}
{Goldreich} P.,  {Tremaine} S.,  1980, \mn@doi [\apj] {10.1086/158356}, \href
  {http://adsabs.harvard.edu/abs/1980ApJ...241..425G} {241, 425}

\bibitem[\protect\citeauthoryear{{Haiman}, {Kocsis}  \& {Menou}}{{Haiman}
  et~al.}{2009}]{2009ApJ...700.1952H}
{Haiman} Z.,  {Kocsis} B.,   {Menou} K.,  2009, \mn@doi [\apj]
  {10.1088/0004-637X/700/2/1952}, \href
  {http://adsabs.harvard.edu/abs/2009ApJ...700.1952H} {700, 1952}

\bibitem[\protect\citeauthoryear{Ivanov, Papaloizou  \& Polnarev}{Ivanov
  et~al.}{1999}]{Ivanov99}
Ivanov P.~B.,  Papaloizou J. C.~B.,   Polnarev A.~G.,  1999, \mn@doi [MNRAS]
  {10.1046/j.1365-8711.1999.02623.x}, 307, 79

\bibitem[\protect\citeauthoryear{{Kelley}, {Blecha}, {Hernquist}  \&
  {Sesana}}{{Kelley} et~al.}{2017}]{Kelley+2017}
{Kelley} L.~Z.,  {Blecha} L.,  {Hernquist} L.,   {Sesana} A.,  2017, preprint,
  \href {http://adsabs.harvard.edu/abs/2017arXiv170202180K} {} (\mn@eprint
  {arXiv} {1702.02180})

\bibitem[\protect\citeauthoryear{{Kocsis} \& {Sesana}}{{Kocsis} \&
  {Sesana}}{2011}]{2011MNRAS.411.1467K}
{Kocsis} B.,  {Sesana} A.,  2011, \mn@doi [\mnras]
  {10.1111/j.1365-2966.2010.17782.x}, \href
  {http://adsabs.harvard.edu/abs/2011MNRAS.411.1467K} {411, 1467}

\bibitem[\protect\citeauthoryear{{Kocsis}, {Haiman}  \& {Loeb}}{{Kocsis}
  et~al.}{2012a}]{Kocsis+2012a}
{Kocsis} B.,  {Haiman} Z.,   {Loeb} A.,  2012a, \mn@doi [\mnras]
  {10.1111/j.1365-2966.2012.22129.x}, 427, 2660

\bibitem[\protect\citeauthoryear{{Kocsis}, {Haiman}  \& {Loeb}}{{Kocsis}
  et~al.}{2012b}]{Kocsis+2012b}
{Kocsis} B.,  {Haiman} Z.,   {Loeb} A.,  2012b, \mn@doi [\mnras]
  {10.1111/j.1365-2966.2012.22118.x}, 427, 2680

\bibitem[\protect\citeauthoryear{{Lin} \& {Papaloizou}}{{Lin} \&
  {Papaloizou}}{1986}]{1986ApJ...309..846L}
{Lin} D.~N.~C.,  {Papaloizou} J.,  1986, \mn@doi [\apj] {10.1086/164653}, \href
  {http://adsabs.harvard.edu/abs/1986ApJ...309..846L} {309, 846}

\bibitem[\protect\citeauthoryear{{Lippai}, {Frei}  \& {Haiman}}{{Lippai}
  et~al.}{2009}]{Lippai+2009}
{Lippai} Z.,  {Frei} Z.,   {Haiman} Z.,  2009, \mn@doi [\apj]
  {10.1088/0004-637X/701/1/360}, \href
  {http://adsabs.harvard.edu/abs/2009ApJ...701..360L} {701, 360}

\bibitem[\protect\citeauthoryear{{MacFadyen} \&
  {Milosavljevi{\'c}}}{{MacFadyen} \&
  {Milosavljevi{\'c}}}{2008}]{2008ApJ...672...83M}
{MacFadyen} A.~I.,  {Milosavljevi{\'c}} M.,  2008, \mn@doi [\apj]
  {10.1086/523869}, \href {http://adsabs.harvard.edu/abs/2008ApJ...672...83M}
  {672, 83}

\bibitem[\protect\citeauthoryear{{Mayer}}{{Mayer}}{2013}]{Mayer:2013:MBHBGasRev}
{Mayer} L.,  2013, \mn@doi [Classical and Quantum Gravity]
  {10.1088/0264-9381/30/24/244008}, \href
  {http://adsabs.harvard.edu/abs/2013CQGra..30x4008M} {30, 244008}

\bibitem[\protect\citeauthoryear{{Miranda}, {Mu{\~n}oz}  \& {Lai}}{{Miranda}
  et~al.}{2017}]{2017MNRAS.466.1170M}
{Miranda} R.,  {Mu{\~n}oz} D.~J.,   {Lai} D.,  2017, \mn@doi [\mnras]
  {10.1093/mnras/stw3189}, \href
  {http://adsabs.harvard.edu/abs/2017MNRAS.466.1170M} {466, 1170}

\bibitem[\protect\citeauthoryear{{Noble}, {Mundim}, {Nakano}, {Krolik},
  {Campanelli}, {Zlochower}  \& {Yunes}}{{Noble}
  et~al.}{2012}]{2012ApJ...755...51N}
{Noble} S.~C.,  {Mundim} B.~C.,  {Nakano} H.,  {Krolik} J.~H.,  {Campanelli}
  M.,  {Zlochower} Y.,   {Yunes} N.,  2012, \mn@doi [\apj]
  {10.1088/0004-637X/755/1/51}, \href
  {http://adsabs.harvard.edu/abs/2012ApJ...755...51N} {755, 51}

\bibitem[\protect\citeauthoryear{{Orosz}, , {Welsh}  \& {et. al}}{{Orosz}
  et~al.}{2012}]{Orosz:2012Sci}
{Orosz} J.~A.,   {Welsh} W.~F.,   {et. al} 2012, \mn@doi [Science]
  {10.1126/science.1228380}, \href
  {http://adsabs.harvard.edu/abs/2012Sci...337.1511O} {337, 1511}

\bibitem[\protect\citeauthoryear{{Paardekooper} \& {Mellema}}{{Paardekooper} \&
  {Mellema}}{2006}]{PM2006}
{Paardekooper} S.-J.,  {Mellema} G.,  2006, \mn@doi [\aap]
  {10.1051/0004-6361:20066304}, \href
  {http://adsabs.harvard.edu/abs/2006A%26A...459L..17P} {459, L17}

\bibitem[\protect\citeauthoryear{{Paczy{\'n}sky} \& {Wiita}}{{Paczy{\'n}sky} \&
  {Wiita}}{1980}]{1980A&A....88...23P}
{Paczy{\'n}sky} B.,  {Wiita} P.~J.,  1980, \aap, \href
  {http://adsabs.harvard.edu/abs/1980A%26A....88...23P} {88, 23}

\bibitem[\protect\citeauthoryear{Peters}{Peters}{1964}]{Peters1964}
Peters P.~C.,  1964, \mn@doi [Physical Review] {10.1103/PhysRev.136.B1224},
  136, 1224

\bibitem[\protect\citeauthoryear{{Rafikov}}{{Rafikov}}{2013}]{Rafikov2012}
{Rafikov} R.~R.,  2013, \mn@doi [\apj] {10.1088/0004-637X/774/2/144}, 774, 144

\bibitem[\protect\citeauthoryear{{Rafikov}}{{Rafikov}}{2016}]{2016ApJ...827..111R}
{Rafikov} R.~R.,  2016, \mn@doi [\apj] {10.3847/0004-637X/827/2/111}, \href
  {http://adsabs.harvard.edu/abs/2016ApJ...827..111R} {827, 111}

\bibitem[\protect\citeauthoryear{{Ragusa}, {Lodato}  \& {Price}}{{Ragusa}
  et~al.}{2016}]{Ragusa+2016}
{Ragusa} E.,  {Lodato} G.,   {Price} D.~J.,  2016, \mn@doi [\mnras]
  {10.1093/mnras/stw1081}, \href
  {http://adsabs.harvard.edu/abs/2016MNRAS.460.1243R} {460, 1243}

\bibitem[\protect\citeauthoryear{{Roedig}, {Sesana}, {Dotti}, {Cuadra},
  {Amaro-Seoane}  \& {Haardt}}{{Roedig} et~al.}{2012}]{2012A&A...545A.127R}
{Roedig} C.,  {Sesana} A.,  {Dotti} M.,  {Cuadra} J.,  {Amaro-Seoane} P.,
  {Haardt} F.,  2012, \mn@doi [\aap] {10.1051/0004-6361/201219986}, \href
  {http://adsabs.harvard.edu/abs/2012A%26A...545A.127R} {545, A127}

\bibitem[\protect\citeauthoryear{{Ryan} \& {MacFadyen}}{{Ryan} \&
  {MacFadyen}}{2017}]{RyanMacFadyen2017}
{Ryan} G.,  {MacFadyen} A.,  2017, \mn@doi [\apj]
  {10.3847/1538-4357/835/2/199}, \href
  {http://adsabs.harvard.edu/abs/2017ApJ...835..199R} {835, 199}

\bibitem[\protect\citeauthoryear{{Sesana}, {Haardt}, {Madau}  \&
  {Volonteri}}{{Sesana} et~al.}{2005}]{Sesana+2005}
{Sesana} A.,  {Haardt} F.,  {Madau} P.,   {Volonteri} M.,  2005, \mn@doi [\apj]
  {10.1086/428492}, \href {http://adsabs.harvard.edu/abs/2005ApJ...623...23S}
  {623, 23}

\bibitem[\protect\citeauthoryear{{Sesana}, {Roedig}, {Reynolds}  \&
  {Dotti}}{{Sesana} et~al.}{2012}]{Sesana+2012}
{Sesana} A.,  {Roedig} C.,  {Reynolds} M.~T.,   {Dotti} M.,  2012, \mn@doi
  [\mnras] {10.1111/j.1365-2966.2011.20097.x}, \href
  {http://adsabs.harvard.edu/abs/2012MNRAS.420..860S} {420, 860}

\bibitem[\protect\citeauthoryear{{Shakura} \& {Sunyaev}}{{Shakura} \&
  {Sunyaev}}{1973}]{1973A&A....24..337S}
{Shakura} N.~I.,  {Sunyaev} R.~A.,  1973, \aap, \href
  {http://adsabs.harvard.edu/abs/1973A%26A....24..337S} {24, 337}

\bibitem[\protect\citeauthoryear{Shi, Krolik, Lubow  \& Hawley}{Shi
  et~al.}{2012}]{0004-637X-749-2-118}
Shi J.-M.,  Krolik J.~H.,  Lubow S.~H.,   Hawley J.~F.,  2012, \apj, 749, 118

\bibitem[\protect\citeauthoryear{{Syer} \& {Clarke}}{{Syer} \&
  {Clarke}}{1995}]{1995MNRAS.277..758S}
{Syer} D.,  {Clarke} C.~J.,  1995, \mn@doi [\mnras] {10.1093/mnras/277.3.758},
  \href {http://adsabs.harvard.edu/abs/1995MNRAS.277..758S} {277, 758}

\bibitem[\protect\citeauthoryear{{Tanaka}, {Menou}  \& {Haiman}}{{Tanaka}
  et~al.}{2012}]{TMH2012}
{Tanaka} T.,  {Menou} K.,   {Haiman} Z.,  2012, \mn@doi [\mnras]
  {10.1111/j.1365-2966.2011.20083.x}, 420, 705

\bibitem[\protect\citeauthoryear{{Volonteri}, {Haardt}  \& {Madau}}{{Volonteri}
  et~al.}{2003}]{Volonteri+2003}
{Volonteri} M.,  {Haardt} F.,   {Madau} P.,  2003, \mn@doi [\apj]
  {10.1086/344675}, \href {http://adsabs.harvard.edu/abs/2003ApJ...582..559V}
  {582, 559}

\bibitem[\protect\citeauthoryear{{Ward}}{{Ward}}{1997}]{1997Icar..126..261W}
{Ward} W.~R.,  1997, \mn@doi [\icarus] {10.1006/icar.1996.5647}, \href
  {http://adsabs.harvard.edu/abs/1997Icar..126..261W} {126, 261}

\makeatother
\end{thebibliography}
\bibliographystyle{mnras}
\label{lastpage}
\end{document}